# Improved Absolute Frequency Measurement of the $^{171}$Yb Optical Lattice Clock at KRISS Relative to the SI Second

Huidong Kim[1], Myoung-Sun Heo[1]*, Won-Kyu Lee[1,2], Chang Yong Park[1,2], Hyun-Gue Hong[1], Sang-Wook Hwang[1], and Dai-Hyuk Yu[1]

[1]*Korea Research Institute of Standards and Science, Daejeon 34113, South Korea*
[2]*University of Science and Technology, Yuseong-gu, Daejeon 34113, South Korea*

*E-mail: hms1005@kriss.re.kr

Abstract

We measured the absolute frequency of the $^1S_0$–$^3P_0$ transition of $^{171}$Yb atom confined in a one-dimensional optical lattice relative to the SI second. The determined frequency was 518 295 836 590 863.38(57) Hz. The uncertainty was reduced by a factor of 14 compared with our previously reported value in 2013 due to the significant improvements in decreasing the systematic uncertainties. This result is expected to contribute to the determination of a new recommended value for the secondary representations of the second.

Optical clocks have made remarkable progress in the last decade, and some of them have already surpassed the best accuracy of Cs microwave clocks by two orders of magnitude.[1-4] Accordingly, a redefinition of the SI second, which is currently based on a microwave transition of Cs atom, has been seriously considered by using optical clocks.[5-7] $^{171}$Yb optical lattice clocks are one of the best developed optical lattice clocks which have an advantage of lower quantum projection noise than single trapped-ion clocks, and are being investigated vigorously by several research groups around the world.[8-16] The frequency of the $^1S_0$–$^3P_0$ transition of $^{171}$Yb has been included in the secondary representation of the SI second by the International Committee for Weights and Measures (CIPM2015)[17] using the reported values of the absolute frequency measurements.[8-11, 18, 19] Since 2015, two more measurement values were reported which will contribute to a new determination of the recommended frequency value in the future.[12, 20]

Since the first absolute frequency measurement of the Yb lattice clock at KRISS (Korea Research Institute of Standards and Science),[10] there have been significant improvements to reduce the systematic uncertainty. The details of the experiment were reported previously[10, 21, 22], thus, they will be described only briefly in this article emphasizing the systematic improvements thereafter. The singlet transition ($^1S_0$–$^1P_1$) at 399 nm was used for the Zeeman slowing and the first stage magneto-optical trap (MOT) with the temperature of about 1 mK. The atoms were further cooled to about 20 μK by the second stage MOT using the triplet transition ($^1S_0$–$^3P_1$) at 556 nm. Next, the atoms were transferred to a vertically-oriented 1D optical lattice trap at the magic wavelength at 759 nm, and were spin-polarized alternatively to either one of the two magnetic sublevels of the ground state ($^1S_0$) by optical pumping using $^1S_0$–$^3P_1$ transition at 556 nm and the bias magnetic field of 0.275 mT.

The optical lattice was formed by a build-up cavity (finesse ~220) to enhance the maximum lattice trap depth ($U_0$) up to ~2500 $E_r$, where $E_r$ is the recoil energy. The polarization of the lattice laser was filtered by a thin-film linear polarizer which was placed between two cavity mirrors as a folding mirror. The extinction ratio was enhanced to more than 1:10,000 by the cavity. The frequency of the lattice laser was stabilized to a fiber frequency comb and monitored with a wavemeter. During transfer of atoms to the lattice, the trap depth was actively stabilized to 880 $E_r$ using an acousto-optic modulator (AOM). Before the clock interrogation, atoms at the higher vibration levels were blown out of the trap using the so-called "energy filtering" method, [22, 23] for which the trap depth was ramped down to 120 $E_r$ in 20 ms and ramped up to 880 $E_r$ in 20 ms.

Just before the clock interrogation, the trap depth was controlled to the target value by a servo loop. Through the sideband spectrum (Fig. 1(a)), the final temperatures of the trapped atoms were estimated to be 4 μK with the trap depth of 163(2) $E_r$. The clock transition ($^1S_0$–$^3P_0$) was interrogated with a clock laser at 578 nm using Rabi spectroscopy with a 40 ms pulse. The clock laser polarization was linear and parallel to the lattice laser polarization. The characteristics of the clock laser was described in Ref. 21. The temperature of the cavity for the clock laser was stabilized to that for zero coefficient of thermal expansion, so that only a linear drift (0.07 Hz/s) due to aging remained. The frequency noise of all the optical fiber transfer routes for the clock laser (to the cavity, to the Yb atom chamber, and to the optical frequency comb) was actively compensated. In particular, the fiber noise in the route to the Yb atom was cancelled at the point after the AOM, which was used for switching, scanning, and steering the clock laser, by coupling the zeroth-order AOM output back to the fiber. This minimized the AOM frequency chirp effect.[25]

After the Rabi interrogation, the normalized population of the excited state was obtained using a repumping laser at 1388 nm ($^3P_0$–$^3D_1$) and a probe laser at 399 nm. Typical normalized clock transition spectrum is shown in Fig. 1(b). The clock laser frequency was locked to the center of the two π-transitions of the clock transition using a digital servo loop and a double-pass AOM. The clock cycle time was about 3.2 s. The overall system was stable and robust such that continuous clock operation of 65,000 s was possible.

To investigate the uncertainty contributions from the systematic effects, we interleaved two different values of the corresponding systematic effect, while the clock laser frequency was locked to the clock transition with an excitation fraction of about 0.5. A typical stability of the interleaved measurement in terms of overlapping Allan deviation is shown in Fig. 2(a). The stability was $7.5\times10^{-15}/\tau^{1/2}$ and the statistical uncertainty reached $1\times10^{-16}$ at 5,800 s.

To evaluate the lattice light shift, we measured the differential shift by interleaving the high (335(3) $E_r$) and low (161(2) $E_r$) lattice depth for each frequency of the lattice light (Fig. 2(b)). Since the frequency of the lattice laser was stabilized to an optical frequency comb referenced to a hydrogen maser[21], we could determine the magic frequency for the electric dipole (E1) polarizability to be 394 798 264.7(84) MHz. The E1 lattice shift was 82(18) mHz considering that the operating condition for the lattice laser was at 394 798 222.9 MHz, $U_0$ = 163(2) $E_r$, and the average vibrational quantum number $\bar{n}$ = 1.1(1). The nonlinear light shift was estimated to be 35(19) mHz, by using the measured values of $U_0$, $\bar{n}$, and the known values of the multipolar polarizability coefficient[20] and the hyperpolarizability coefficient[9].

To evaluate the collisional frequency shift at the atomic density $\rho$, we measured the clock laser frequency locked to the clock transition, alternating the high and low atomic density (Fig. 2(c)). The atom number and accordingly the atomic density were varied by changing the Zeeman slowing laser power. The density shift could be fitted to $-55(12)$ mHz $\times \rho/\rho_0$, where $\rho_0 = 1\times10^9/\text{cm}^3$. The operating condition for the atomic density was $7(4)\times10^8/\text{cm}^3$, thus, the shift was estimated to be -38(21) mHz.

The shifts by the linear Zeeman effect and the lattice vector polarizability were removed in the process of the digital servo lock to the clock transition using the mean value of the two π-transition resonance frequencies ($m_F = \pm1/2 \to m_F' = \pm1/2$). The quadratic Zeeman shift was estimated to be -529(76) mHz using the known coefficient[9] and the magnetic field value of 0.275 mT measured by the Zeeman splitting.

The black body radiations (BBRs) from finite-temperature surroundings shift the clock transition. We applied a model used in Ref. 14 and 26 which considered two contributions, one from the outside of the main chamber (direct line of sight) and the other from the inner surfaces in thermal equilibrium. In our case, the first contribution came from the room through 14 chamber windows (eight 1.5-inch windows and six 1-inch windows made of fused silica) 110 mm away from atoms and the direct radiation from a 1-inch heated window 700 mm away from atoms. The BBR frequency shift from the room (22.6(10) °C) through fourteen windows was assumed to be negligible because the transmittance of a fused silica window is very low at the wavelength larger than 4 μm, however, we took the uncertainty of 23 mHz, estimating conservatively the uncertainty of the transmittance to be 0.2. The room temperature was measured by two calibrated PT100 sensors. The BBR shift from the heated window at 192(10) °C was -0.66(6) mHz. The second contribution came from the inner surfaces of chamber wall and windows. We assumed an isotropic room temperature radiation, diffuse reflections from the inner surfaces and spatially homogeneous isotropic energy density.[14,26] The emissivity of the Aquadag[27]-coated chamber wall was 0.8(1) and that of the windows was assumed to be 0.8(2). Temperatures of inner surfaces were nearly the same within 1 °C (22.6(10) °C) measured by six calibrated PT100 sensors, except that of the Zeeman slower nipple at 35(10) °C. By equating total energy flux into and out of the chamber sphere[14, 26], we found the BBR shift from the inner surfaces to be -1.206(16) Hz.

Stray electric fields around atoms can induce a Stark shift in the clock transition[12, 28]. The vacuum chamber made of stainless steel effectively shielded the electric fields from the outside. Electric charges on glass windows 110 mm away from atoms have negligible effect because their electric fields were attenuated by a factor of $>10^4$. Near the trap center, there were two identical aspheric lenses (outer diameter : 25 mm, radius of curvature : 19.7 mm) made of fused silica located symmetrically at 34.6 mm away from atoms inside the vacuum chamber for the fluorescence collection. Electric charges present on these lenses could decay via copper tubes holding these lenses to the electrical ground. The capacitance between the two lenses was estimated to be 0.05 pF. Assuming a high volume resistivity of $10^{18}$ Ωm for high-purity fused silica, we estimated a decay time constant of about 90 days. Since these lenses have been inside the vacuum chamber for more than 1.7 years, a Stark shift from the remaining charges was below $1\times10^{-17}$ assuming maximum allowed electric field under vacuum ($4\times10^4$ V/m).[29]

The systematic uncertainty budget is summarized in Table I with the total fractional uncertainty of $1.8\times10^{-16}$. The frequency shifts due to the background gas, the AOM chirp, the fiber noise cancellation, the line pulling, and the lattice tunneling were estimated to be well below $1\times10^{-17}$ and could be neglected in this uncertainty evaluation.

The absolute frequency of the clock transition ($^1S_0$–$^3P_0$) of $^{171}$Yb atom referenced to the SI second was measured during ten consecutive days from MJD 57734 to MJD 57743. Durations of each measurement were between 4,000 ~ 65,000 s. We measured the frequency of the clock laser by using an optical frequency comb referenced to a hydrogen maser. A dead-time-free frequency counter in pi-mode with the gate time of 1 s was used. This measured frequency was compared with the interpolated values of the clock laser frequency simultaneously measured by the Yb atoms per cycle time of 3.2 s. The result is shown in Fig. 2(d). The clock operation covers 34% of the total measurement duration. The cycle slips in the frequency counting were excluded by comparing the simultaneous measurement data from three independent counters. The relative frequency offset of the hydrogen maser from TAI (International Atomic Time) for the ten-day measurement interval was calculated to be 9.458(65)×10$^{-14}$ by using BIPM Circular-T no. 348[30]) and the time comparison data between the hydrogen maser and UTC(KRIS). The dead-time uncertainty in calculating the hydrogen maser offset was estimated to be 5.0×10$^{-16}$ from the noise characteristics of the hydrogen maser[31]). The frequency offset and the dead-time uncertainty of TAI from the SI second (TT) were -12.4(22)×10$^{-16}$ and 5.2×10$^{-16}$ using Circular-T no. 348 during the ten-day-long measurement interval. The height of Yb atoms in optical lattice from the earth ellipsoid (WGS84) was measured to be 74.60(3) m by a GPS antenna data and orthometric height levelling. The geoid undulation was obtained by the Earth Gravitational Model (EGM2008)[32]). Taking the bias by the tidal effect as an additional uncertainty, the gravitational shift was conservatively estimated to be 81.4(33) ×10$^{-16}$. As summarized in Table I, the absolute frequency of the clock transition relative to the SI second was determined to be 518 295 836 590 863.38(57) Hz whose uncertainty was dominated by the link to TAI. The result is compared with previous measurements in Fig. 3. The measured absolute frequency value in this work agrees well with the CIPM recommendation in 2015[17]) and the previously reported values[8-12, 18-20]) with the uncertainty decreased by 14-fold compared with our previous measurement. We expect this value will contribute to the determination of a new recommended value for the secondary representations of the second by CIPM and eventually to a future redefinition of the SI second.

**Acknowledgments**

The authors thank Sung-Hoon Yang, Young Kyu Lee, Jong Koo Lee, and Chang Bok Lee for providing the frequency data of the H-maser and UTC(KRIS). This work was supported partly by the Korea Research Institute of Standards and Science under the project "Research on Time and Space Measurements," Grant No. 17011005, and also partly by the R&D Convergence Program of NST (National Research Council of Science and Technology) of Republic of Korea (Grant No. CAP-15-08-KRISS).

32) http://earth-info.nga.mil/GandG/wgs84/gravitymod/egm2008/egm08_wgs84.html

**Figure Captions**

**Fig. 1.** (a) Sideband spectrum of the clock transition using $1\times10^4$ atoms. The blue dots are experimental data and the blue line is combination of fits to the carrier and sidebands.[24] (b) Typical normalized spectrum of the single magnetic component of the clock transition with 40 ms pulse at the lattice depth of 163 $E_r$. The experimental data (black dots) agrees well with the sinc function (black line) with the Fourier limited linewidth of 20 Hz.

**Fig. 2.** (a) Instability of a typical interleaved measurement in terms of Allan deviation. (b) Measured differential lattice light shift by interleaving the trap depth of 335 $E_r$ and 161 $E_r$ as a function of the lattice laser frequency. The line is a linear fit to the data corrected for the collisional shift and the non-linear lattice shift. (c) Measured density shift by interleaving between the high and low atomic densities is fitted to a straight line (red line) with error (shadow). (d) Absolute frequency measurement data of the $^{171}$Yb lattice clock relative to the SI second from MJD 57734 to MJD 57743. The red line is the weighted mean value and the shadow is the total uncertainty.

**Fig. 3.** Absolute frequency values of $^{171}$Yb clock transition measured by different laboratories with relevant references. Black squares; relative to the SI second (TT), blue diamonds; deduced form $^{171}$Yb/$^{87}$Sr frequency ratio measurement, red circles; relative to Cs fountain standards. The green solid line and the shaded region represent the CIPM recommended value and its uncertainty, respectively; $f$(CIPM2015) = 518 295 836 590 864.0(10) Hz[17].

**Table I.** Uncertainty budget for the absolute frequency measurement of the $^{171}$Yb clock transition.

| Effect | Relative Shift $\times 10^{-16}$ | Relative Uncertainty $\times 10^{-16}$ |
|---|---|---|
| Lattice ac Stark (scalar) | 1.6 | 0.4 |
| Nonlinear lattice shift | 0.7 | 0.4 |
| Density shift | -0.7 | 0.4 |
| Blackbody radiation | -23.3 | 0.6 |
| Second-order Zeeman | -10.2 | 1.5 |
| Probe light | 0.2 | 0.04 |
| AOM phase chirp | - | 0.03 |
| Servo error | - | 0.7 |
| Static Stark shift | - | 0.1 |
| Yb total | -31.8 | 1.8 |
| Statistics | - | 4.0 |
| Gravitational red shift | 81.4 | 0.3 |
| H-maser dead-time | - | 5.0 |
| H-maser - TAI | 945.8 | 6.5 |
| TAI dead-time | - | 5.2 |
| TAI - TT | -12.4 | 2.2 |
| Total | 983.0 | 10.9 |

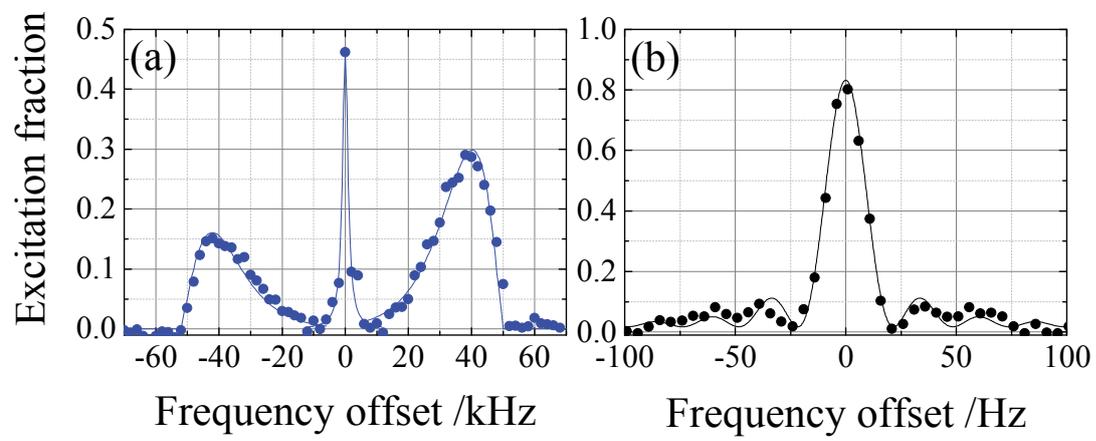

Fig.1.

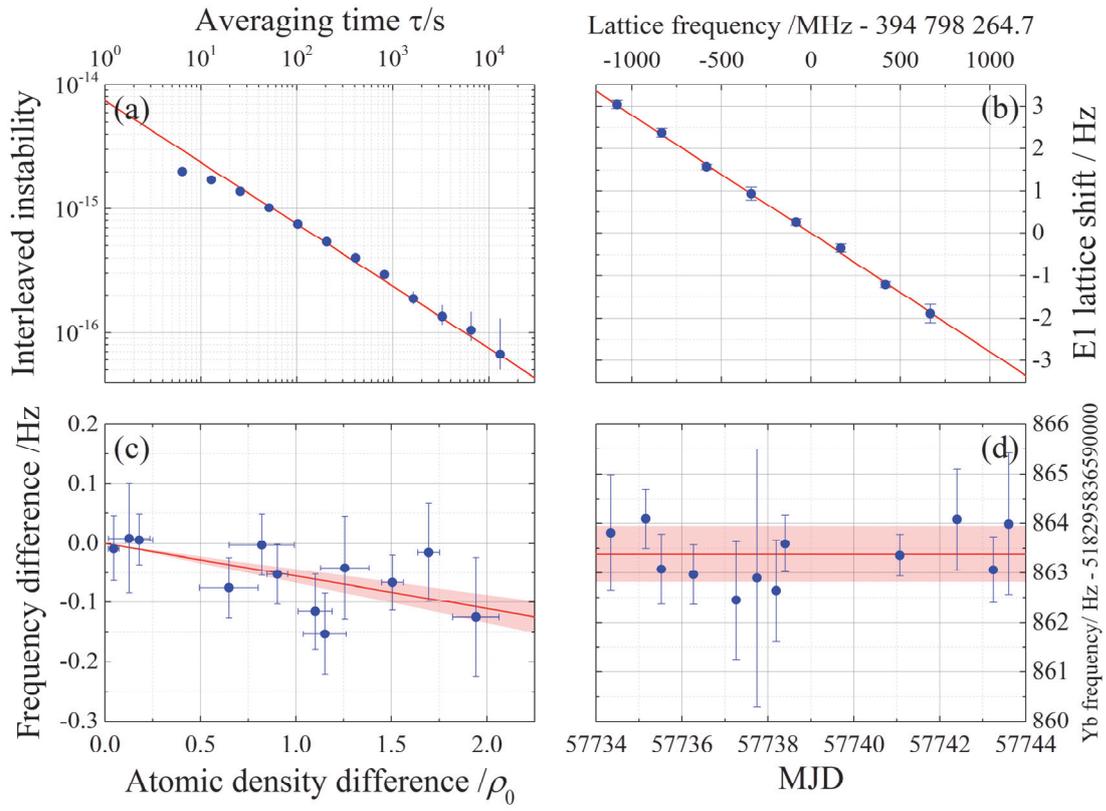

Fig.2.

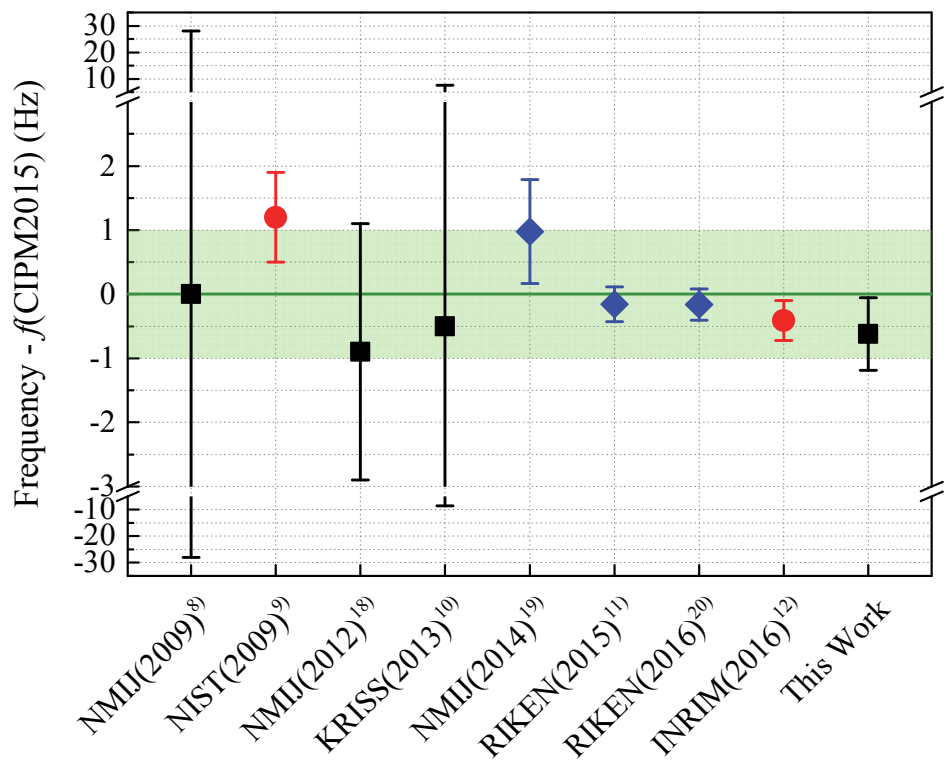

Fig. 3.